\documentclass[12pt,showpacs,amsmath,amssymb,aps,floats,floatfix,nofootinbib]{revtex4-1}
\usepackage{hyperref}
\usepackage{epsfig}
\usepackage{dcolumn}
\usepackage{bm}
\usepackage{amsmath}
\usepackage{amsfonts}
\usepackage{amssymb}
\usepackage{graphicx,subfigure}
\usepackage{color}
\usepackage{latexsym}
\usepackage{slashed} 
\usepackage{indentfirst}
\usepackage{mathrsfs}
\usepackage{multirow}
\usepackage{epsfig,subfigure,psfrag}

\usepackage{axodraw}
\graphicspath{{fig/}} 

\def\fb{\mathrm{fb}} 
\def\ifb{\mathrm{fb}^{-1}} 
\def\TeV{\mathrm{TeV}} 
\def\GeV{\mathrm{GeV}} 

\makeatother

\allowdisplaybreaks 

\begin{document}

\title{The 750 GeV diphoton resonance as a singlet scalar in an extra dimensional model}

\author{Chengfeng Cai$^1$}
\author{Zhao-Huan Yu$^2$}
\author{Hong-Hao Zhang$^1$\footnote{Email: zhh98@mail.sysu.edu.cn}}

\affiliation{$^1$School of Physics, Sun Yat-Sen University, Guangzhou 510275, China}
\affiliation{$^2$ARC Centre of Excellence for Particle Physics at the Terascale,
School of Physics, The~University of Melbourne, Victoria 3010, Australia}

\begin{abstract}

We interpret the 750~GeV diphoton excess recently found in the 13~TeV LHC data as a singlet scalar in an extra dimensional model, where one extra dimension is introduced.
In the model, the scalar couples to multiple vector-like fermions, which are just the KK modes of SM fermions.
Mediated by the loops of these vector-like fermions, the $\phi$ effective couplings to gluons and photons can be significantly large.
Therefore, it is quite easy to obtain an observed cross section for the diphoton excess.
We also calculate the cross sections for other decay channels of $\phi$, and find that this interpretation can evade the bounds from the 8~TeV LHC data.

\end{abstract}

\maketitle

\section{Introduction}
\label{sec:intro}

Recently, the ATLAS collaboration reported an excess at $\sim 750~\GeV$ in the diphoton invariant mass distribution, based on the 13~TeV LHC data with an integrated luminosity of $3.2~\ifb$~\cite{ATLAS:2015diphoton}.
Assuming this excess is due to a resonance, the local (global) signal significance is $3.9\sigma$ ($2.3\sigma$).
A very preliminary fit showed that the resonance has a broad width of $\sim 45~\GeV$.
However, since the statistics of the current data is not quite sufficient, the measurement of the width should be treated as a hint rather than a final conclusion.
On the other hand, a similar excess at $\sim 760~\GeV$ have also been found by the CMS collaboration based on a data set of $2.6~\ifb$, but the local (global) significance is lower and just 2.6$\sigma$ (1.2$\sigma$)~\cite{CMS:2015dxe}.

These exciting results have stimulated lots of theoretical interpretations~\cite{Harigaya:2015ezk,Mambrini:2015wyu,Backovic:2015fnp,Angelescu:2015uiz,Nakai:2015ptz,Knapen:2015dap,Buttazzo:2015txu,Pilaftsis:2015ycr,Franceschini:2015kwy,DiChiara:2015vdm,McDermott:2015sck,Ellis:2015oso,Low:2015qep,Bellazzini:2015nxw,Gupta:2015zzs,Petersson:2015mkr,Molinaro:2015cwg,Dutta:2015wqh,Cao:2015pto,Matsuzaki:2015che,Kobakhidze:2015ldh,Martinez:2015kmn,Cox:2015ckc,Becirevic:2015fmu,No:2015bsn,Demidov:2015zqn,Chao:2015ttq,Fichet:2015vvy,Curtin:2015jcv,Bian:2015kjt,Chakrabortty:2015hff,Ahmed:2015uqt,Agrawal:2015dbf,Csaki:2015vek,Falkowski:2015swt,Aloni:2015mxa,Bai:2015nbs,Ghosh:2015apa,Gabrielli:2015dhk,Benbrik:2015fyz,Kim:2015ron,Alves:2015jgx,Megias:2015ory,Carpenter:2015ucu,Bernon:2015abk,Chao:2015nsm,Arun:2015ubr,Han:2015cty,Chang:2015bzc,Ringwald:2015dsf,Chakraborty:2015jvs,Ding:2015rxx,Han:2015dlp,Hatanaka:2015qjo,Luo:2015yio,Chang:2015sdy,Bardhan:2015hcr,Feng:2015wil,Antipin:2015kgh,Wang:2015kuj,Cao:2015twy,Huang:2015evq,Dhuria:2015ufo,Bi:2015uqd,Kim:2015ksf,Berthier:2015vbb,Cline:2015msi,Chala:2015cev,Kulkarni:2015gzu,Barducci:2015gtd,Boucenna:2015pav,Murphy:2015kag,Hernandez:2015ywg,Dey:2015bur,Pelaggi:2015knk,Dev:2015isx,Huang:2015rkj,Chabab:2015nel,Moretti:2015pbj,Patel:2015ulo,Badziak:2015zez,Chakraborty:2015gyj,Cao:2015xjz,Cvetic:2015vit,Gu:2015lxj,Allanach:2015ixl,Davoudiasl:2015cuo,Das:2015enc,Cheung:2015cug,Liu:2015yec,Zhang:2015uuo,Casas:2015blx,Hall:2015xds}.
Although some works suggested this excess may be due to a spin-2 particle~\cite{Arun:2015ubr,Han:2015cty}
or a spin-1 particle\footnote{In this case, the decay products of the resonance should involve, besides two photons, at least one extra particle. Otherwise it would contradict the Landau-Yang theorem~\cite{Landau:1948kw,Yang:1950rg}.}~\cite{Bernon:2015abk,Chala:2015cev,Liu:2015yec},
most of these works interpreted it as a new scalar ($\phi$) beyond the standard model (SM).
New scalars can be naturally introduced from the Higgs sector in many SM extensions.
However, since the observed cross section for $pp\to\phi\to\gamma\gamma$ is $\sim \mathcal{O}(10)~\fb$, ordinary two-Higgs-doublet and supersymmetric models could not give such a large production cross section without further extensions~\cite{Angelescu:2015uiz,Buttazzo:2015txu,DiChiara:2015vdm}.

As LHC is a $pp$ collider, parton distribution functions determine that $gg$ fusion happens much more often than $q\bar{q}$ annihilation.
Besides, $\phi$ couplings to quarks are usually proportional to quark masses in many models.
Therefore, $gg$ fusion should be the dominant process for $pp\to\phi$ production.
In order to increase the $pp\to\phi\to\gamma\gamma$ production rate, the  $\phi$ couplings to $gg$ and $\gamma\gamma$, which are generally induced by loop processes, need to be significantly enhanced.
This can be achieved by introducing multiple electrically charged and colored vector-like fermions coupled to $\phi$~\cite{Angelescu:2015uiz,Dutta:2015wqh,Benbrik:2015fyz,Boucenna:2015pav,Gu:2015lxj,Zhang:2015uuo}.
The vector-like feature is particularly appealing for avoiding gauge anomalies.

In this work, we give a reasonable origin for such vector-like fermions: they are KK modes of SM fermions in an SM extension where just one compactified extra dimension is introduced.
We assume that there is a 5D CP-even singlet scalar field $\Phi$
that couples to all 5D fermion fields and generates their bulk mass terms by obtaining a nonzero vacuum expectation value (VEV).
Once nonzero bulk masses are generated, the profiles of the zero modes will be exponential functions so that the zero modes are localized at either end of the interval of the 5th dimension.
The localization of fermions is especially attractive because it can give an explanation to the fermion mass hierarchy problem~\cite{ArkaniHamed:1999dc,Kaplan:2001ga,Kitano:2003cn,Fujimoto:2012wv,Cai:2015jla}.
Consequently, although the 5D fermions are born as vector-like fields, their zero modes will be chiral after imposing Dirichlet boundary conditions and become the usual SM fermions,
while higher KK modes remain vector-like.

Here this localizing feature is caused by the scalar VEV, thus $\Phi$ is called the localizer~\cite{Kaplan:2001ga}.
We assume the excitation around the VEV of the zero mode of $\Phi$ is the observed 750~GeV scalar $\phi$. $\phi$ automatically couples to the KK modes of the fermions, which are vector-like and charged under the electroweak gauge symmetry.
Therefore $\phi$ can be produced through the gluon-gluon fusion process induced by the KK quark loops and then decays into two photons induced by the KK quark and KK charged lepton loops.

Rather than covering complicated setups, we fully simplify the model and just focus on how to explain the diphoton excess.
Therefore, the gluon-gluon fusion process is assumed to be the only source for $pp\to\phi$ production.
After production, $\phi$ decays into $gg$, $\gamma\gamma$, $ZZ$, $\gamma Z$, and $W^+W^-$ through loops, as well as $hh$ at the tree level,
which would be suppressed if the mixing between $\phi$ and the SM Higgs is small.
We will calculate the cross sections for these processes and demonstrate that the model is consistent with all existed bounds.

The paper is organised as follows.
In Sec.~\ref{sec:model} we briefly introduce the model.
In Sec.~\ref{sec:mixing} we discuss the KK modes of $\phi$ and the mixing between $\phi$ and the Higgs.
Sec.~\ref{sec:analy} presents our interpretation to the 750~GeV diphoton excess and show that it is consistent with 8~TeV LHC bounds.
Sec.~\ref{sec:concl} gives the conclusion.

\section{The model}
\label{sec:model}

We discuss a quite simplified model, which is similar to the one in Ref.~\cite{Kaplan:2001ga}.
We assume all fields corresponding to SM particles are living in a flat 5D space-time. Rather than considering a orbifold structure and orbifold boundary conditions~\cite{Georgi:2000wb}, we simply adopt a compactified interval as the fifth dimension. The usual 4D coordinates and the extra dimensional coordinate are denoted as $x^\mu$ and $y\in[0,\pi R]$, respectively. As in universal extra dimensions~\cite{Appelquist:2000nn}, gauge fields have boundary conditions that $A_y=0$ and $\partial_y A_\mu=0$ at $y=0$ and $\pi R$, and hence their zero modes have flat profiles $f_\mathrm{gauge}(y)=1/\sqrt{\pi R}$. The Higgs field distributes in the 5D space-time with a flat VEV, and its zero mode also has a flat profile.

Apart from the Higgs, a new singlet scalar $\Phi$ is introduced. We assume the VEV of $\Phi$ is flat, in order to generate constant bulk masses for all fermions. After solving the equation of motion with proper boundary conditions on the fermions, we can obtain chiral zero-mode fermions, which play the role of ordinary SM fermions.
The profiles of these zero modes are localized at either end of the interval and exponentially spread into the 5th dimension.

To be precise, we consider the action describing a 4-component 5D fermion $\Psi$ coupled to $\Phi$ as
\begin{eqnarray}\label{action}
S=\int d^5x[\bar{\Psi}(i\Gamma^MD_M-\tilde{y}_f\Phi)\Psi]=\int d^5x[\bar{\Psi}(i\Gamma^MD_M-\tilde{y}_f\langle\Phi\rangle-\tilde{y}_f\tilde{\Phi}))\Psi],
\end{eqnarray}
where $\Gamma^M$ is the gamma matrices with the fifth matrix defined as $\Gamma^5=i\gamma^5$.
The covariant operator $D_M=\partial_M-i\tilde{g}A_M$, where $A_M$ is a gauge field and $\tilde{g}$ is a gauge coupling with a mass dimension of $-1/2$.
After $\Phi$ develops a VEV $v_\phi$, the fermion acquires a bulk mass
\begin{equation}
M_f=\tilde{y}_f\langle\Phi\rangle=y_fv_\phi,
\end{equation}
where $y_f=\tilde{y}_f/\sqrt{\pi R}$ is a dimensionless Yukawa coupling.
$\tilde{\Phi}$ is the excitation around the VEV.

The kinetic and bulk mass terms in the action~\eqref{action} give the following equation of motion for a free fermion field:
\begin{eqnarray}\label{eom}
\begin{pmatrix}\partial_y-M_f&i\sigma^\mu\partial_\mu\\i\bar{\sigma}^\mu\partial_\mu&-\partial_y-M_f\end{pmatrix}\begin{pmatrix}\Psi_L\\\Psi_R\end{pmatrix}=0.
\end{eqnarray}
Dirichlet boundary conditions will be imposed on either the left-handed or the right-handed component of $\Psi$.
These two components can be decomposed in modes:
\begin{equation}
\Psi_L=\sum_{n=0}^\infty\chi_a^{(n)}(x)f^{(n)}(y),\quad
\Psi_R=\sum_{n=0}^\infty\xi^{(n)\dag\dot{a}}(x)g^{(n)}(y).
\end{equation}
Substituting the mode decomposition into Eq.~\eqref{eom}, we obtain
\begin{equation}
(-\partial_y+M_f)f^{(n)}(y)=m_n g^{(n)}(y),\quad
(\partial_y+M_f)g^{(n)}(y)=m_n f^{(n)}(y).
\end{equation}
Thus the 4D field of each mode satisfies the Dirac equation.

The profile of a zero mode with a zero mass ($m_0=0$) is $f^{(0)}(y)\propto e^{M_fy}$ and $g^{(0)}(y)\propto e^{-M_fy}$.  Then if we impose the Dirichlet boundary condition that $\Psi_R=0$ ($\Psi_L=0$) at $y=0$ and $\pi R$, we will have $g^{(0)}(y)=0$ ($f^{(0)}(y)=0$), which means that the zero mode is chiral as desired for SM fermions.
Note that the zero mode here does not have mass, which can be acquired from the ordinary Higgs mechanism as we will see below.

On the other hand, the KK modes ($n\geq1$) are vector-like, i.e., their left-handed and right-handed components transform as the same under a gauge symmetry. The profiles of these KK modes can be exactly solved. If the boundary condition is $\Psi_R=0$ at $y=0$ and $\pi R$, then
\begin{eqnarray}
f^{(n)}(y)&=&\sqrt{\frac{2/\pi R}{M_f^2+n^2/R^2}}\left(\frac{n}{R}\cos\frac{ny}{R}+M_f\sin\frac{ny}{R}\right), \label{LH:f}\\
g^{(n)}(y)&=&\sqrt{\frac{2}{\pi R}}\sin\frac{ny}{R}.\label{LH:g}
\end{eqnarray}
The mass of the $n$-mode is given by $m_n^2=M_f^2+n^2M_{\mathrm{KK}}^2$  ($n\geq 1$), where $M_{\mathrm{KK}}\equiv R^{-1}$. In this model, we assume that the 5D fermion fields for $SU(2)_L$ doublet fermions have this kind of boundary conditions, so that their zero modes are left-handed.
For the 5D fermion fields for $SU(2)_L$ singlet fermions, we impose the boundary condition that $\Psi_L=0$ at $y=0$ and $\pi R$, and hence their zero modes are right-handed.

Substituting the mode expansion of $\Psi$ into the action, we can obtain the couplings between $\tilde{\Phi}$ and each mode of the fermion. After integrating out the 5th dimension coordinate, the effective 4D action of the Yukawa interactions corresponding to a fermion with a left-handed zero mode can be expressed as
\begin{eqnarray}\label{phiFF0}
S_{\phi \bar{F}F}&=&\int d^4x\left\{-y_f\sum_{n=1}^\infty\frac{M_f}{\sqrt{M_f^2+n^2M_{\mathrm{KK}}^2}}\phi(x)\bar{F}^{(n)}(x)F^{(n)}(x)\right.\nonumber\\
&&-\frac{2y_f}{\sqrt{\pi}}\sqrt{\frac{M_f}{e^{2\pi M_f/M_{\mathrm{KK}}}-1}}\sum_{n=1}^\infty\frac{nM_{\mathrm{KK}}^{-3/2}[1-\cos(n\pi)e^{\pi M_f/M_{\mathrm{KK}}}]}{M_f^2+n^2M_{\mathrm{KK}}^2}\phi(x)[\bar{F}^{(n)}(x)F^{(0)}_L(x)+\mathrm{h.c.}]\nonumber\\
&&\left.-\frac{y_f}{\pi}\sum_{m\pm n=\mathrm{odd}}\frac{4mn}{n^2-m^2}\frac{M_{\mathrm{KK}}}{\sqrt{M_f^2+m^2M_{\mathrm{KK}}^2}}\phi(x)[\bar{F}^{(n)}(x)F^{(m)}_L(x)+\mathrm{h.c.}]\right\},
\end{eqnarray}
where the 4D scalar field $\phi(x)$ corresponds to the zero mode of $\tilde{\Phi}$, and $F^{(n)}(x)$ is the 4D $n$-mode Dirac fermion. For $n\geq 1$, $F^{(n)}(x)$ is vector-like.
Note that there is no $\phi\bar{F}^{(0)}F^{(0)}$ coupling in the action~\eqref{phiFF0}.
Below we will see that the coupling in the first line will contribute to the LHC production and decay of $\phi$ through loop processes. If the bulk mass $M_f$ vanishes, the $\phi\bar{F}^{(n)}F^{(n)}$ couplings would vanish as well, even in the case that $y_f$ is non-zero and $M_f$ is tuned to be zero by adding another constant bulk mass to compensate $\tilde{y}_f\left<\Phi\right>$.
Therefore, a non-zero bulk mass is necessary for loop production and decay of $\phi$ in this model.

\begin{figure}[!htbp]
\centering
\includegraphics[width=0.35\textwidth]{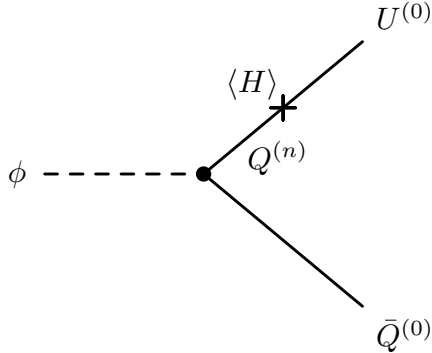}
\caption{Forbidden decay $\phi\to U^{(0)}\bar{Q}^{(0)}$.}
\label{fig:phi:U0Q0}
\end{figure}

For simplicity, we assume that $\tilde{y}_q$ and $\tilde{y}_l$ are universal for all quarks and leptons, respectively.
For the $SU(2)_L$ fermions, we assume that the singlet Yukawa couplings to $\Phi$ are opposite to the corresponding doublet Yukawa couplings.
That is to say, the Yukawa terms for the quark doublet $Q$ and the lepton doublet $L$ can be given by $-\tilde{y}_q\Phi\bar{Q}Q - \tilde{y}_l\Phi\bar{L}L$,
while those for the singlets $U$, $D$, and $E$ are $+\tilde{y}_q\Phi\bar{U}U +\tilde{y}_q\Phi\bar{D}D +\tilde{y}_l\Phi\bar{E}E$.
Then the profiles of the KK modes of a doublet has the forms of Eqs.~\eqref{LH:f} and \eqref{LH:g}, while the profiles of the KK modes of a singlet are given by
\begin{eqnarray}
f_s^{(n)}(y)&=-&\sqrt{\frac{2}{\pi R}}\sin\frac{ny}{R},\\
g_s^{(n)}(y)&=&\sqrt{\frac{2/\pi R}{M_f^2+n^2/R^2}}(\frac{n}{R}\cos\frac{ny}{R}+M_f\sin\frac{ny}{R}).
\end{eqnarray}
With this special setup, $g_s^{(n)}(y)$ is the same as $f^{(n)}(y)$ in Eq.~\eqref{LH:f}. If there is Yukawa couplings of the Higgs field to the doublets and singlets, e.g., $-\tilde{y}_u\bar{Q}i\sigma_2H^\ast U+\mathrm{h.c.}$, due to the orthogonality between $g_s^{(n)}(y)$ and $f^{(0)}(y)$, in the effective 4D action the Higgs boson would only connect the same modes:
\begin{equation}
-y_u \left[\bar{Q}^{(0)}(x) i\sigma_2 H^{(0)\ast}(x)U^{(0)}(x)+\sum_{n=1}^{\infty}\bar{Q}^{(n)}(x)i\sigma_2 H^{(0)\ast}(x)U^{(n)}(x)\right]+\mathrm{h.c.}.
\end{equation}
Consequently, a mixing mass term between the zero mode of a singlet (doublet) and a KK mode of a doublet (singlet) cannot be generated by the Higgs VEV due to this orthogonality. It also kills any decay process of $\phi$ like Fig.~\ref{fig:phi:U0Q0}, therefore $\phi$ can neither be produced from $q\bar{q}$ nor decay to quarks and leptons.

Now we discuss how the fermion KK modes couple with the gauge fields.
The zero mode of a gauge field has a flat profile, while its KK modes have profiles as
\begin{eqnarray}
f_\mathrm{gauge}^{(n)}=\sqrt{\frac{2}{\pi R}}\cos\frac{ny}{R}.
\end{eqnarray}
Masses of these KK modes are $nM_{\mathrm{KK}}$, if we neglect the contributions for $W$ and $Z$ from the Higgs VEV.
From the flat profile of the zero-mode gluon and the orthogonality between different KK modes, we can obtain the interactions between the zero-mode gluon and quarks:
\begin{eqnarray}
\mathcal{L}_{g\bar{Q}Q}&=&g_s\sum_{n=1}^{\infty}(\bar{Q}^{(n)}\slashed{G}Q^{(n)}+\bar{U}^{(n)}\slashed{G}U^{(n)}+\bar{D}^{(n)}\slashed{G}D^{(n)})\nonumber\\
&&+g_s(\bar{Q}^{(0)}_L\slashed{G}Q^{(0)}_L+\bar{U}_R^{(0)}\slashed{G}U_R^{(0)}+\bar{D}_R^{(0)}\slashed{G}D_R^{(0)}).
\end{eqnarray}
The interaction between the photon and fermions are similar except that the gauge coupling is replaced by the electric charge.

\begin{figure}[!htbp]
\centering
\subfigure[~$gg\to\phi$]
{\includegraphics[width=0.35\textwidth]{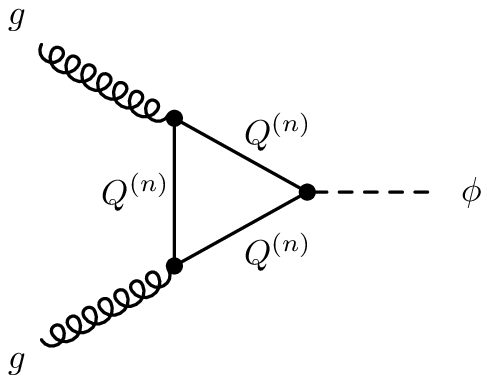}}
\hspace*{2em}
\subfigure[~$\phi\to\gamma\gamma$]
{\includegraphics[width=0.35\textwidth]{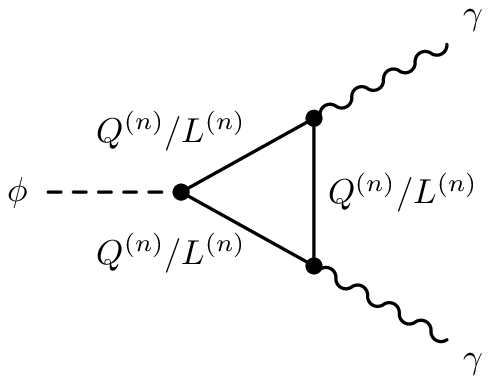}}
\caption{Feynman diagrams for gluon-gluon fusion to $\phi$~(a)
and $\phi\to \gamma\gamma$ decay~(b) through the loops of $Q^{(n)}$ and $L^{(n)}$. There are also analogous diagrams through the loops of $U^{(n)}$, $D^{(n)}$, and $E^{(n)}$.}
\label{fig:gg:phi:2gamma}
\end{figure}

The Feynman diagrams for $gg\to\phi$ and $\phi\to\gamma\gamma$ are shown in Fig.~\ref{fig:gg:phi:2gamma}.
The effective operators of $\phi GG$ and $\phi AA$ couplings can be obtained by integrate out the fermion loops.
This is similar to the SM Higgs case.
Adopting the parametrization in Ref.~\cite{Carmi:2012in}, the effective operators are
\begin{eqnarray}
\mathcal{L} \supset \kappa_g\frac{\alpha_s}{12\pi v_\phi}\phi G_{\mu\nu}^aG^{a\mu\nu}+\kappa_\gamma\frac{\alpha}{\pi v_\phi}\phi A_{\mu\nu}A^{\mu\nu},
\end{eqnarray}
where the factor $\kappa_g$ and $\kappa_\gamma$ come from loop integration and are mainly contributed by fermion loops in our model:
\begin{eqnarray}
\kappa_g&=&\sum_{f}2C(r_f)\kappa_fA_f(\tau_f),\\
\kappa_\gamma&=&\sum_{f}\frac{N_c(r_f)Q_f^2}{6}\kappa_fA_f(\tau_f),
\end{eqnarray}
where $C(r_f)$ and $N_c(r_f)$ are the index and the dimension of the $SU(3)_C$ representation where the fermion lives in, respectively.
$Q_f$ is the electric charge of the fermion.
The loop function $A_f(\tau_f)$ is defined as
\begin{eqnarray}
A_f(\tau_f)=\frac{3}{2\tau_f^2}[(\tau_f-1)\arcsin^2\sqrt{\tau_f}+\tau_f]
\end{eqnarray}
with $\tau_f\equiv m_\phi^2/4m_f^2\leq1$.
The factor $\kappa_f$ corresponds to the $\phi\bar{F}^{(n)}F^{(n)}$ coupling
\begin{eqnarray}
\mathcal{L}\supset-\kappa_f^{(n)}\frac{m_n}{v_\phi}\phi \bar{F}^{(n)}F^{(n)}.
\end{eqnarray}
From the action~\eqref{phiFF0}, we can read off the factor as
\begin{equation}
\kappa_f^{(n)}=\frac{y_f^2v_\phi^2}{m_n^2}.
\end{equation}
Combining all the results above, one can calculate the gluon-gluon fusion production cross section for $\phi$ and its diphoton partial decay width.
Since there are 3 generations of fermions with different KK modes, the production rate for $pp\to\phi\to\gamma\gamma$ would be significantly increased.

\section{The KK modes of $\Phi$ and scalar mixing}
\label{sec:mixing}

In this section, we discuss more details about the singlet scalar $\Phi$ to justify the model.
According to the equation of motion of the scalar in a compactified extra dimension, the $n$-mode mass of $\Phi$ is given by
\begin{eqnarray}
m_{\phi^{(n)}}^2=n^2M_{\mathrm{KK}}^2-M^2,
\end{eqnarray}
where $M$ comes from the quadratic term of $\Phi$.
In order to give bulk masses for fermions, the quadratic and quartic terms of $\Phi$ in the potential must be appropriate to make the zero mode $\phi$ develops a nonzero VEV.

However, if $M_{\mathrm{KK}}<M$, the 1-mode of $\Phi$ would also develop a nonzero VEV. To circumvent this complication, we assume that $M_{\mathrm{KK}}>M$. Moreover, for avoiding possible experimental constraints, it would be necessary to make the 1-mode heavier than the 0-mode, i.e., $m_{\phi^(1)}^2=M_{\mathrm{KK}}^2-M^2>m_\phi^2$.
If $M_{\mathrm{KK}}=1$~TeV, this relation will be satisfied for $M\lesssim 660~\GeV$ given $m_\phi = 750~\GeV$.
As $\phi^{(1)}$ have similar interactions as $\phi$, if it is just slightly heavier than $\phi$, it would be promising to discover it in the following LHC runs.

Since the profiles of different modes are orthogonal to each other, there is no mixing term between $\phi$ and $\phi^{(n)}$.
Nonetheless, the mixing between $\phi$ and the 0-mode of the Higgs field $H$ is inevitable. Although more data are required to increase the statistics, current LHC results suggest that the Higgs couplings to SM particles are quite consistent with the standard model.
Thus it would be more safe to demand a small mixing.
The 4D potential with $\phi$ and $H$ involved is
\begin{eqnarray}
V(H,\phi)&=&-\frac{\mu^2}{2}(v+h)^2+\frac{1}{4}\lambda(v+h)^4-\frac{M^2}{2}(v_\phi+\phi)^2+\frac{\lambda_\phi}{4!}(v_\phi+\phi)^4\nonumber\\
&&+\frac{\lambda_{\phi h}}{2}(v+h)^2(v_\phi+\phi)^2.
\end{eqnarray}
The VEVs are obtained by minimizing the potential:
\begin{equation}
v=\sqrt{\frac{\mu^2-\lambda_{\phi h}v_\phi^2/2}{\lambda}},\quad
v_\phi=\sqrt{\frac{6(M^2-\lambda_{\phi h}v^2/2)}{\lambda_\phi}}.
\end{equation}
According to the first expression, for natural values $v_\phi\sim 1~\TeV$ and $\lambda_{\phi h}v_\phi^2/2 \sim \mathcal{O}((100~\GeV)^2)$, a small mixing coupling $\lambda_{h\phi}$ of $\mathcal{O}(10^{-2})$ is needed.
The mass matrix for $h$ and $\phi$ is
\begin{eqnarray}
\begin{pmatrix}m_h^2&\lambda_{\phi h}vv_\phi\\\lambda_{\phi h}vv_\phi&m_\phi^2\end{pmatrix},
\end{eqnarray}
where $m_h^2=2\lambda v^2$ and $m_\phi^2=\lambda_\phi v_\phi^2/3$.
The physical masses are given by
\begin{eqnarray}
m_{h_{1,2}}^2=\frac{1}{2}\left[m_h^2+m_\phi^2\mp\sqrt{(m_\phi^2-m_h^2)^2+4\lambda_{h\phi}v^2v_\phi^2}\right].
\end{eqnarray}
If $m_\phi^2\gg2\lambda_{h\phi}vv_\phi$, they are essentially decoupled, and we have $m_{h_1}^2\approx m_h^2$ and $m_{h_2}^2\approx m_\phi^2$.

Parametrizing the mixing matrix as
\begin{eqnarray}
\begin{pmatrix}h_1\\h_2\end{pmatrix}=\begin{pmatrix}\cos\alpha&-\sin\alpha\\\sin\alpha&\cos\alpha\end{pmatrix}\begin{pmatrix}h\\ \phi\end{pmatrix},
\end{eqnarray}
we can find that
\begin{eqnarray}
\tan2\alpha=\frac{2\lambda_{h\phi}vv_\phi}{m_\phi^2-m_h^2}.
\end{eqnarray}
in the small $\alpha$ limit,
\begin{eqnarray}
\sin\alpha\approx\frac{\lambda_{h\phi}vv_\phi}{m_\phi^2-m_h^2}.
\end{eqnarray}
As $m_\phi\approx750~\GeV\approx 6m_h$, we have
\begin{equation*}
m_\phi^2-m_h^2\approx\frac{35}{6}m_hm_\phi=\frac{35}{6}\sqrt{\frac{2}{3}\lambda\lambda_\phi}vv_\phi.
\end{equation*}
Thus the small mixing condition corresponds to
\begin{eqnarray}
 \sin\alpha\approx\frac{6\sqrt{3}\lambda_{h\phi}}{35\sqrt{2\lambda\lambda_{h\phi}}}\ll1.
\end{eqnarray}
For instance, $\sin\alpha\approx0.1$ needs $\lambda_{h\phi}<0.17\sqrt{\lambda_\phi}$.

On the other hand, the value of $\lambda_{h\phi}$ also affects the tree level decay $\phi\to hh$, whose partial width is approximately given by
\begin{equation}
\Gamma(\phi\to hh) \approx \frac{\sqrt{2}m_\phi}{16\pi}\frac{\lambda_{h\phi}^2}{\lambda_\phi^2}.
\end{equation}
If $\lambda_{h\phi}\sim\lambda_\phi$, we would have $\Gamma(\phi\to hh)\sim 0.03 m_\phi$.
Then this channel would be dominant and give a broad total width.
However, it would be easily excluded by the 8~TeV LHC result $\sigma(pp\to\phi\to hh)<39~\mathrm{fb}$~\cite{ATLAS:2014rxa}.
For these reasons, below we will just consider the small mixing case and fix $\lambda_{h\phi} = 0.01$.

\section{Interpretation to the 750~GeV diphoton resonance}
\label{sec:analy}

In the model, the effective operators for $\phi$ couplings to gluons and photons can be explicitly expressed as
\begin{eqnarray}
\mathcal{L}_{\phi gg}&=&\frac{\alpha_sv_\phi}{\pi}\left[\sum_{n=1}^{n_*}\frac{y_q^2 A_f(\tau_n)}{y_q^2v_\phi^2+n^2M_{\mathrm{KK}}^2}\right]\phi G_{\mu\nu}^aG^{a\mu\nu},\\
\mathcal{L}_{\phi\gamma\gamma}&=&\frac{\alpha}{\pi}\left[\sum_{n=1}^{n_*}\frac{5}{3}\frac{y_q^2v_\phi A_f(\tau_q^{(n)})}{y_q^2v_\phi^2+n^2M_{\mathrm{KK}}^2}
+\sum_{n=1}^{n_*}\frac{y_l^2v_\phi A_f(\tau_l^{(n)})}{y_l^2v_\phi^2+n^2M_{\mathrm{KK}}^2}\right]\phi A_{\mu\nu}A^{\mu\nu},
\end{eqnarray}
where we have included all vector-like fermion loops, and $n_*$ is the maximum of $n$ we consider.
Note that the model we discuss in this paper is actually an effective model, and the perturbative unitarity could be violated for a large $n_*$.
According to the unitarity arguments in Ref.~\cite{Chivukula:2003kq}, we have $n<3$ for this model.
Thus we adopt $n_*=2$ in the following calculation to obtain an optimized enhancement for the $pp\to\phi\to\gamma\gamma$ production.

The free parameters in the model are $v_\phi$, $M_\mathrm{KK}$, $y_q$, and $y_l$.
The Yukawa couplings  $y_q$ and $y_l$ should not to be too large to remain perturbative.
The $\phi$ decay channels involved are $\phi\to gg$, $\gamma\gamma$, $hh$, $Z\gamma$, $ZZ$, and $W^+W^-$. The $hh$ channel is the only tree-level decay process and the rest are generated by vector-like fermion loops.
We numerically calculate the partial widths for the loop-induced channels using the code FeynCalc~\cite{Mertig:1990an} and LoopTools~\cite{Hahn:1998yk}.

Under the narrow width approximation, the cross section for $pp\to\phi\to X_1X_2$ can be computed by
\begin{equation}
\sigma(pp\to\phi\to X_1X_2)=\left(\frac{k_{gg}}{0.1}\right)^2 \sigma_{\mathrm{ref}}(pp\to\phi) \mathrm{Br}(\phi\to X_1X_2)
\end{equation}
with
\begin{equation}
k_{gg}=\frac{1~\TeV}{M_\mathrm{KK}}\frac{\alpha_s v_\phi}{\pi M_\mathrm{KK}}\sum_{n=1}^{n_*}\frac{y_q^2 A_f(\tau_n)}{y_q^2v_\phi^2/M_{\mathrm{KK}}^2+n^2},
\end{equation}
where $\mathrm{Br}(\phi\to X_1X_2)$ is the branching ratio of the $\phi\to X_1X_2$ decay.
We calculate $\sigma_{\mathrm{ref}}(pp\to\phi)$ using \texttt{Madgraph~5}~\cite{Alwall:2014hca} and \texttt{FeynRules}~\cite{Alloul:2013bka},
and obtain $\sigma_{\mathrm{ref}}(pp\to\phi)=16.725~(3.783)$~pb at the 13~TeV (8~TeV) LHC.

In the following, we investigate which values of the parameters can give a cross section of $pp\to\phi\to \gamma\gamma$ consistent with observation.
In this model, the $\phi\to gg$ partial width is much larger than that of any other channel. Therefore we can roughly estimate the branching ratio of $\phi\to \gamma\gamma$ as
\begin{equation}
\mathrm{Br}(\phi\to \gamma\gamma)\approx\frac{\Gamma(\phi\to \gamma\gamma)}{\Gamma(\phi\to gg)}=\frac{8\alpha^2}{9\alpha_s^2}
\end{equation}
for $y_l=y_q$. Thus the cross section $\sigma(pp\to\phi\to \gamma\gamma)$ at 13~TeV is
\begin{eqnarray}
\sigma(pp\to\phi\to \gamma\gamma)&\approx&
\frac{800\alpha^2}{9\pi^2}\sigma_{\mathrm{ref}}(pp\to\phi)\frac{v_\phi^2}{M_{\mathrm{KK}}^2}\left|\sum_{n=1}^{n_*}\frac{y_q^2 A_f(\tau_n)}{y_q^2v_\phi^2/M_{\mathrm{KK}}^2+n^2}\right|^2\nonumber\\
&\approx& 9.2~\fb \cdot\frac{v_\phi^2}{M_{\mathrm{KK}}^2}\left|\sum_{n=1}^{n_*}\frac{y_q^2 A_f(\tau_n)}{y_q^2v_\phi^2/M_{\mathrm{KK}}^2+n^2}\right|^2.
\end{eqnarray}
The diphoton excess signal at the 13~TeV LHC corresponds to $\sigma(pp\to\phi\to \gamma\gamma)=5-20~\fb$.
Apparently, if $v_\phi\sim M_{\mathrm{KK}}\sim1$~TeV and $y_{q}\sim1$, $\sigma(pp\to\phi\to \gamma\gamma)$ at 13~TeV would be around 9~fb, which is favored by current data.

On the other hand, there are some constraints for resonances from LHC Run~1 data.
Relevant 95\% C.L. bounds at the 8~TeV LHC include~\cite{Aad:2015kna,Khachatryan:2015cwa,Aad:2015agg,Aad:2014fha,CMS:2015neg,Aad:2014aqa,ATLAS:2014rxa}:
\begin{eqnarray}
&&\sigma(pp\to\phi\to ZZ)<12~\mathrm{fb},\\
&&\sigma(pp\to\phi\to W^+W^-)<40~\mathrm{fb},\\
&&\sigma(pp\to\phi\to Z\gamma)<4~\mathrm{fb},\\
&&\sigma(pp\to\phi\to jj)<2.5~\mathrm{pb},\\
&&\sigma(pp\to\phi\to hh)<39~\mathrm{fb}.
\end{eqnarray}
In the model, $\phi$ could decay into SM quarks and leptons through the mixing with the Higgs boson. However, since we have chosen a very small mixing parameter, these decay channels can be neglected.
Therefore, the $pp\to\phi\to jj$ process principally comes from the $\phi\to gg$ decay, and the $pp\to\phi\to l^+l^-$ process is irrelevant to the phenomenology here.

After calculation, it turns out that the $pp\to\phi\to Z\gamma$, $jj$, and $hh$ searches at the 8~TeV LHC can hardly constrain the relevant region of the parameter space, and we will not plot their bounds in the following figures.
Fig.~\ref{fig:vphiyql} shows the contours of $\sigma_{\gamma\gamma}\equiv\sigma(pp\to\phi\to \gamma\gamma)$ in the $v_\phi$-$y_{q,l}$ plane for $M_{\mathrm{KK}}=1~\TeV$ assuming $y_l=y_q$.
We can see that $\sigma_{\gamma\gamma}=5-20~\fb$ corresponds to a large region that does not excluded by the 8~TeV LHC searches for the $pp\to\phi\to ZZ$ and $W^+W^-$ processes.
This means that our interpretation to the diphoton excess is consistent with the 8~TeV LHC data,
although a large $y_{q,l}\gtrsim 1$ is demanded to give a sufficient large $\sigma_{\gamma\gamma}$.
A smaller $v_\phi$ would decrease the masses of vector-like fermions in loops, and hence increase the signal.

\begin{figure}[!htbp]
\centering
\includegraphics[width=0.6\textwidth]{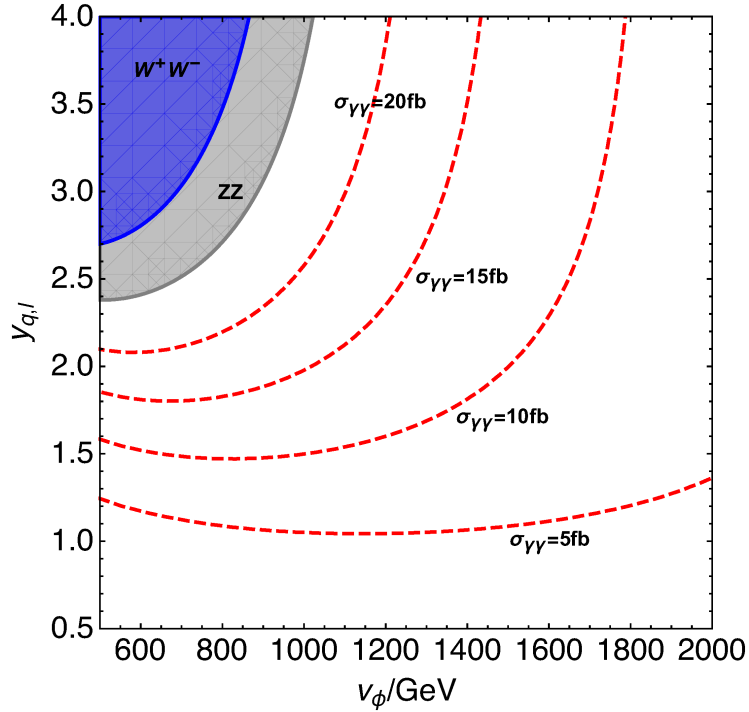}
\caption{Contours of $\sigma_{\gamma\gamma}$ (red dashed lines) in the $v_\phi$-$y_{q,l}$ plane for $M_{\mathrm{KK}}=1~\TeV$ assuming $y_l=y_q$.
The blue (gray) region is excluded at 95\% C.L. by the $pp\to\phi\to W^+W^-~(ZZ)$ search at the 8~TeV LHC.}
\label{fig:vphiyql}
\end{figure}

In order to investigate how large $M_{\mathrm{KK}}$ could be, we fix $v_\phi=1~\TeV$ and demonstrate in Fig.~\ref{fig:Mkkyql} the contours of $\sigma_{\gamma\gamma}$ in the $M_{\mathrm{KK}}$-$y_{q,l}$ plane assuming $y_l=y_q$. We find that if we can tolerate $y_{q,l}$ as large as 5.3, $M_{\mathrm{KK}}$ can reach up to $5~\TeV$ for an desired diphoton signal. On the other hand, if we can just tolerate $y_{q,l}\sim 2$, then $M_{\mathrm{KK}}$ is bounded to below 2~TeV.
We also plot the contours of the $\phi$ total decay width $\Gamma_\mathrm{tot}$ in Fig.~\ref{fig:Mkkyql}.
We find that the predicted $\Gamma_\mathrm{tot}$ could reach up to $\sim 1~\GeV$, which is still smaller that the favored value $45~\GeV$ from the preliminary ATLAS analysis.
If such a broad width persists in the follow-up experiments, extra decay channels might be needed to increase the total width.
For instance, $\phi$ may decay into a pair of DM particles and this can be still consistent the 8~TeV bounds~\cite{Mambrini:2015wyu,Bi:2015uqd}.

\begin{figure}[!htbp]
\centering
\includegraphics[width=0.6\textwidth]{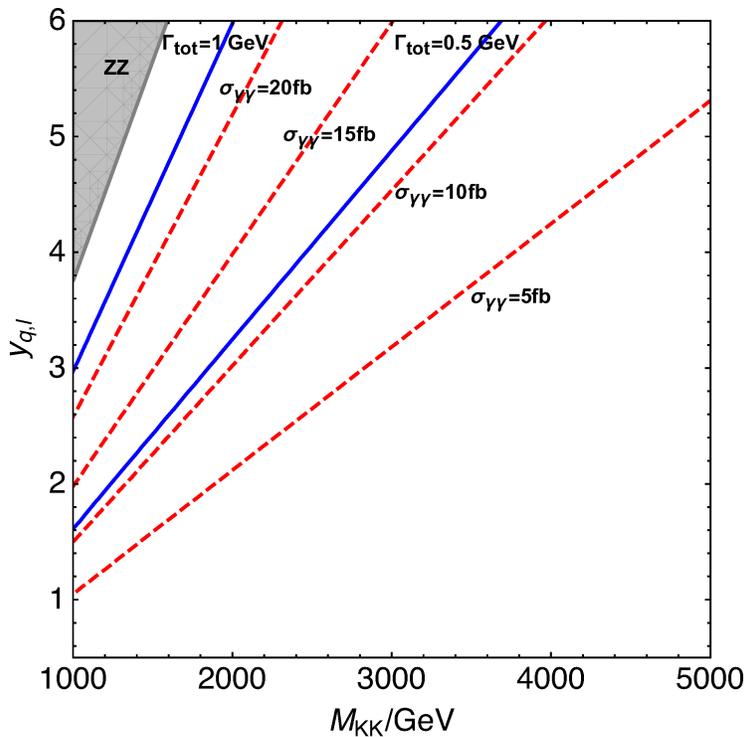}
\caption{Contours of $\sigma_{\gamma\gamma}$ (red dashed lines) and the total width of $\phi$ (blue solid lines) in the $M_{\mathrm{KK}}$-$y_{q,l}$ plane for $v_\phi=1~\TeV$ assuming $y_l=y_q$.
The gray region is excluded at 95\% C.L. by the $pp\to\phi\to ZZ$ search at the 8~TeV LHC.}
\label{fig:Mkkyql}
\end{figure}

Finally, we study how $\sigma_{\gamma\gamma}$ depends on $y_q$ and $y_l$, as presented in Fig.~\ref{fig:yqyl}, where $M_{\mathrm{KK}}=v_\phi=1~\TeV$ is fixed.
Since vector-like quark KK modes both enter the $\phi gg$ and $\phi \gamma\gamma$ effective couplings while vector-like charged lepton KK modes only contribute to the $\phi \gamma\gamma$ effective coupling,
$\sigma_{\gamma\gamma}$ is more sensitive to $y_q$.
As $y_q$ decreases, $y_l$ should be dramatically increased to enhance the $\gamma\gamma$ branching ratio for compensating the signal.
Although $y_l$ may be even allowed to vanish, $y_q$ should be at least $\sim 0.5$ for explaining the diphoton excess.

 \begin{figure}[!htbp]
\centering
\includegraphics[width=0.6\textwidth]{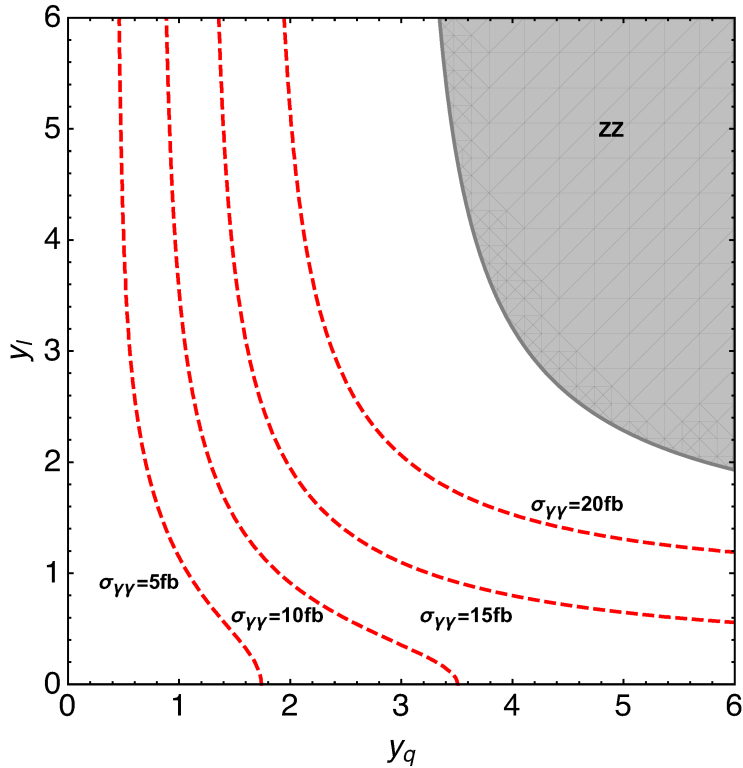}
\caption{Contours of $\sigma_{\gamma\gamma}$ (red dashed lines) in the $y_q$-$y_l$ plane for $M_{\mathrm{KK}}=v_\phi=1~\TeV$.
The gray region is excluded at 95\% C.L. by the $pp\to\phi\to ZZ$ search at the 8~TeV LHC.}
\label{fig:yqyl}
\end{figure}

\section{Conclusions and discussions}
\label{sec:concl}

The 750~GeV diphoton excess recently found in the 13~TeV LHC data have drawn great attention of the high energy physics community.
In this work, we interpret it as a singlet scalar $\phi$ in an extra dimensional model, where just one compactified extra dimension is introduced.

We assume there is a 5D singlet scalar field $\Phi$ coupled to multiple vector-like 5D fermions through Yukawa interactions.
After $\Phi$ gets a VEV, its zero-mode excitation around the VEV becomes the observed scalar $\phi$, and the 5D fermions acquire bulk mass terms, which localize the zero modes of these fermions.
By imposing appropriate boundary conditions, the fermion zero modes become chiral and play the roles of ordinary SM fermions, while the KK modes remain vector-like.
$\phi$ can mix with the SM Higgs boson. However, after considering the LHC measurement of the Higgs couplings as well as the natural choices of model parameters, a small mixing case is favored.

The vector-like fermion KK modes have the same gauge quantum numbers as the corresponding zero modes.
As $\phi$ couples them, they can induce significantly large $\phi gg$ and $\phi \gamma\gamma$ effective couplings through loop diagrams.
Consequently, this model can easily give an observed cross section of $\mathcal{O}(10)~\fb$ for the diphoton excess without contradict the 8~TeV LHC constraints.

Since the decay channels $\phi \to gg$, $ZZ$, $Z\gamma$, and $W^+W^-$ always exist along with the diphoton channel in this model, follow-up experimental searches for $\phi$ through these final states would be crucial to test the model.
On the other hand, if the vector-like quark KK modes have masses of $\sim\mathcal{O}(1)~\TeV$, they could be accessible at the 13~TeV and 14~TeV LHC. Once produced, they can decay into $\phi$ and the corresponding zero-mode quarks and may lead to distinguishable signatures.
If they are heavier, say $\sim\mathcal{O}(10)~\TeV$, it would become a task for future higher energy hadron colliders, such as SppC and FCC-hh.


\acknowledgments

This work is supported by the National Natural Science Foundation of China (NSFC)
under Grant Nos. 11375277, 11410301005, and 11005163, the Fundamental Research
Funds for the Central Universities, and the Sun Yat-Sen University Science Foundation.
ZHY is supported by the Australian Research Council.

\bibliographystyle{JHEP}
\bibliography{ref}

\end{document}